\documentclass[pre,twocolumn,twoside,byrevtex,superscriptaddress,floatfix]{revtex4-1}

\usepackage{graphicx,epsfig,verbatim,enumerate}
\usepackage{amssymb,amsmath}
\usepackage{ifthen,tikz}

\usepackage{hyperref}
\hypersetup{
        colorlinks = true,
        allcolors = blue
}

\usepackage{longtable}

\usepackage{mathtools}

\newboolean{twocolswitch}
\newboolean{normalrefs}

\newcommand{\sindex}[1]{}
\newcommand{\nindex}[1]{}

\newcommand{\etal}{\textit{et al.}}
\newcommand{\www}[1]{\url{#1}}

\usepackage{lettrine}

\usepackage{color}

\newcommand{\havgfn}{h_{\rm avg}}

\setboolean{twocolswitch}{true}
\setboolean{normalrefs}{true}

\begin{document}

\title{\protect
Reply to Garcia et al.: Common mistakes in measuring frequency
dependent word characteristics
}

\author{
\firstname{Peter Sheridan}
\surname{Dodds}
}
\email{peter.dodds@uvm.edu}

\affiliation{
  Computational Story Lab,
  Vermont Advanced Computing Core,
  \& the Department of Mathematics and Statistics, University of Vermont,
  Burlington,
  VT, 05401
}
\affiliation{
  Vermont Complex Systems Center,
  University of Vermont,
  Burlington,
  VT, 05401
}

\author{
\firstname{Eric M.}
\surname{Clark}
}

\affiliation{
  Computational Story Lab,
  Vermont Advanced Computing Core,
  \& the Department of Mathematics and Statistics, University of Vermont,
  Burlington,
  VT, 05401
}
\affiliation{
  Vermont Complex Systems Center,
  University of Vermont,
  Burlington,
  VT, 05401
}

\author{
\firstname{Suma}
\surname{Desu}
}

\affiliation{
  Center for Computational Engineering,
  Massachusetts Institute of Technology,
  Cambridge,
  MA, 02139
}

\author{
\firstname{Morgan R.}
\surname{Frank}
}

\affiliation{
  Center for Computational Engineering,
  Massachusetts Institute of Technology,
  Cambridge,
  MA, 02139
}

\author{
\firstname{Andrew J.}
\surname{Reagan}
}

\affiliation{
  Computational Story Lab,
  Vermont Advanced Computing Core,
  \& the Department of Mathematics and Statistics, University of Vermont,
  Burlington,
  VT, 05401
}
\affiliation{
  Vermont Complex Systems Center,
  University of Vermont,
  Burlington,
  VT, 05401
}

\author{
\firstname{Jake Ryland}
\surname{Williams}
}

\affiliation{
  Computational Story Lab,
  Vermont Advanced Computing Core,
  \& the Department of Mathematics and Statistics, University of Vermont,
  Burlington,
  VT, 05401
}
\affiliation{
  Vermont Complex Systems Center,
  University of Vermont,
  Burlington,
  VT, 05401
}

\author{
\firstname{Lewis}
\surname{Mitchell}
}
\affiliation{
  School of Mathematical Sciences ,
  North Terrace Campus,
  The University of Adelaide,
  SA 5005, Australia
}

\author{
\firstname{Kameron Decker}
\surname{Harris}
}

\affiliation{
  Applied Mathematics, 
  University of Washington,
  Lewis Hall \#202, 
  Box 353925, 
  Seattle, 
  WA, 98195.
}

\author{
\firstname{Isabel M.}
\surname{Kloumann}
}

\affiliation{
  Center for Applied Mathematics,
  Cornell University,
  Ithaca,
  NY, 14853.
}

\author{
\firstname{James P.}
\surname{Bagrow}
}

\affiliation{
  Computational Story Lab,
  Vermont Advanced Computing Core,
  \& the Department of Mathematics and Statistics, University of Vermont,
  Burlington,
  VT, 05401
}
\affiliation{
  Vermont Complex Systems Center,
  University of Vermont,
  Burlington,
  VT, 05401
}

\author{
\firstname{Karine}
\surname{Megerdoomian}
}

\affiliation{
  The MITRE Corporation,
  7525 Colshire Drive,
  McLean, 
  VA, 22102
}

\author{
\firstname{Matthew T.}
\surname{McMahon}
}

\affiliation{
  The MITRE Corporation,
  7525 Colshire Drive,
  McLean, 
  VA, 22102
}

\author{
\firstname{Brian F.}
\surname{Tivnan}
}
\email{btivnan@mitre.org}

\affiliation{
  The MITRE Corporation,
  7525 Colshire Drive,
  McLean, 
  VA, 22102
}

\affiliation{
  Vermont Complex Systems Center,
  University of Vermont,
  Burlington,
  VT, 05401
}

\author{
\firstname{Christopher M.}
\surname{Danforth}
}
\email{chris.danforth@uvm.edu}

\affiliation{
  Computational Story Lab,
  Vermont Advanced Computing Core,
  \& the Department of Mathematics and Statistics, University of Vermont,
  Burlington,
  VT, 05401
}
\affiliation{
  Vermont Complex Systems Center,
  University of Vermont,
  Burlington,
  VT, 05401
}

\date{\today}

\begin{abstract}
  \protect
  We demonstrate that the concerns expressed by Garcia \etal\ 
are misplaced, due to 
(1) a misreading of our findings in~\cite{dodds2015a_onlineappendices};
(2) a widespread failure to examine and present words in support of 
asserted summary quantities based on word usage frequencies;
and
(3) a range of misconceptions about word usage frequency, word rank,
and expert-constructed word lists.
In particular, we show that the English component of our study compares well 
statistically with two related surveys, 
that no survey design influence is apparent,
and that estimates of measurement error do not explain the
positivity biases reported in our work and that of others.
We further demonstrate that for the frequency dependence of positivity---of which we 
explored the nuances in great detail in~\cite{dodds2015a_onlineappendices}---Garcia \etal\ did not 
perform a reanalysis of our data---they instead carried out
an analysis of a different, statistically improper data set
and introduced a nonlinearity before performing linear regression.

\end{abstract}

\maketitle

Note: The present manuscript is an elaboration of our short reply letter~\cite{dodds2015b}.

\section{Function words in the LIWC data set are not emotionally
  neutral}

We first address Garcia \etal's concerns about our 
online survey~\cite{garcia2015a}, which they suggest induced
a positivity bias in respondents' answers.

Garcia \etal\ claim that a set 
of function words in the 
LIWC (Language Inquiry and Word Count)
data set~\cite{pennebaker2007a} 
show a wide spectrum of average 
happiness with positive skew (their Fig 1A) when, according to their
interpretation, these words should exhibit a Dirac delta function
located at neutral ($\havgfn$=5 on a 1 to 9 scale).
We expose and address two fundamental errors.

First, function words in the LIWC data set are simply not emotionally neutral.
The LIWC data set annotates 4487 words and stems
on a wide range of dimensions~\cite{pennebaker2007a}.
We find a total of 421 words and 48 stems are 
coded as function words with 450 matches in our data set when using stems.
Of these, only 7 are indicated as emotional (5 positive, 2 negative)
which appears to support Garcia \etal's interpretation.
However, a straightforward reading of the LIWC list of function words
reveals that these words readily bear emotional weight as exemplified by 
``greatest'' and ``worst''.
We present some of the most extremely and most neutrally rated
LIWC function words in Tab.~\ref{tab:mhl-reply.liwc-function}.

More generally, ``Not looking at the words'' and 
``Not showing the words'' are pervasive issues with 
word- and phrase-based summary statistics for texts.
We should be able to see how specific 
words contribute to summary statistics for texts
to provide
(1) an assurance the measure is performing as intended,
and
(2) insight into the text itself.
For example, all sentiment scoring algorithms based on words
and phrases must be able to plainly show why one text
is more positive through changes in word frequency,
such as through the word shifts we have developed for
both print~\cite{dodds2009b,dodds2011e,dodds2015a}
and as interactive, online visualizations~\cite{hedonometer2015a}.
Elsewhere, in studying the Google Books corpus, 
we have produced analogous word shifts for 
the Jensen-Shannon divergence~\cite{pechenick2015a}.
We exhort other researchers to produce similar word shifts (and not just
word clouds), and to question work with no such counterpart.

Second, 
as we discuss in detail in Sec.~\ref{sec:mhl-liwc.freqdepend} below,
no statement about biases can be made about
sets of words chosen without frequency of usage incorporated.
Any given set of words may have a positive, neutral, or negative 
bias, but we must know how they are chosen before being able
to generalize (as we have done thoroughly in~\cite{dodds2015a_onlineappendices}).
Because we have no guarantee that the expert-generated LIWC function words
are exhaustive 
and because we are merging words of highly variable
usage frequency,
a finding of an average positive bias for LIWC function words 
is meaningless, 
regardless of their transparent capacity for being non-neutral.

\begin{table}
  \centering
  \begin{tabular}{rr}
  \textbf{High} & $\havgfn$ \\
  \hline
billion & 7.56 \\
million & 7.38 \\
couple & 7.30 \\
millions & 7.26 \\
greatest & 7.26 \\
rest & 7.18 \\
best & 7.18 \\
equality & 7.08 \\
unique & 6.98 \\
plenty & 6.98 \\
truly & 6.86 \\
hopefully & 6.84 \\
first & 6.82 \\
plus & 6.76 \\
well & 6.68 \\
greater & 6.68 \\
highly & 6.60 \\
me & 6.58 \\
done & 6.54 \\
extra & 6.52 \\
infinite & 6.44 \\
simply & 6.42 \\
equally & 6.40 \\
sixteen & 6.39 \\
we & 6.38 \\
soon & 6.34 \\
  \end{tabular}
  \quad
  \begin{tabular}{rr}
  \textbf{Neutral} &  $\havgfn$ \\
  \hline
    been & 5.04  \\
    other & 5.04  \\
    into & 5.04  \\
    theyre & 5.04  \\
    it & 5.02  \\
    some & 5.02  \\
    where & 5.02  \\
    themselves & 5.02  \\
    im & 5.02  \\
    quarterly & 5.02  \\
    ive & 5.02  \\
    because & 5.00  \\
    whereas & 5.00  \\
    id & 5.00  \\
    til & 5.00  \\
    the & 4.98  \\
    to & 4.98  \\
    by & 4.98  \\
    or & 4.98  \\
    part & 4.98  \\
    rather & 4.98  \\
    its & 4.96  \\
    when & 4.96  \\
    perhaps & 4.96  \\
    yall & 4.96  \\
    of & 4.94  \\
 \end{tabular}
 \quad
 \begin{tabular}{rr}
  \textbf{Low} &  $\havgfn$  \\
  \hline
   wouldnt & 3.86 \\
   not & 3.86 \\
   shouldn't & 3.84 \\
   none & 3.84 \\
   haven't & 3.82 \\
   wouldn't & 3.78 \\
   fewer & 3.72 \\
   lacking & 3.71 \\
   won't & 3.70 \\
   wasnt & 3.70 \\
   dont & 3.70 \\
   don't & 3.70 \\
   down & 3.66 \\
   nobody & 3.64 \\
   doesn't & 3.62 \\
   couldnt & 3.58 \\
   without & 3.54 \\
   no & 3.48 \\
   cant & 3.48 \\
    zero & 3.44  \\
    against & 3.40  \\
    never & 3.34  \\
    cannot & 3.32  \\
    lack & 3.16  \\
    negative & 2.42  \\
    worst & 2.10  \\
 \end{tabular}
  \caption{
    Three subsets of 450
    LIWC function words with high, neutral, and low 
    average happiness scores
    from our labMT study~\cite{dodds2011e,dodds2015a_onlineappendices} 
    (stems provide more
    matches than those found by Garcia et al.).
    Each word's score is the average rating for 50 participants (scale is 1 to 9
    with 1 = most negative, 5 = neutral, and 9 = most positive).
    Function words may carry emotional weight and cannot be presumed
    to be neutral.
  }
  \label{tab:mhl-reply.liwc-function}
\end{table}

Emotional words in LIWC provide another case in point.
Around 20\% of the LIWC data set (907 words and stems)
are denoted as having positive affect (160 words and 247 stems) or negative affect (151
words and 349 stems).
While stems complicate word counts, the LIWC data set
clearly does not show evidence of a positivity bias.
Because the LIWC data set is expert-curated and meant to be general, 
it does not fit any natural
corpora with respect to usage frequency (i.e., LIWC words 
constitute an unsystematic sampling).
Word lists meant to accurately reflect statistical properties of
language must be built directly from the most frequently
used words of well defined corpora---a point we will return
to several times in this reply.
An earlier expert-curated word list, the smaller ANEW data
set~\cite{bradley1999a}, similarly fails in these respects,
showing a fairly flat distribution across average happiness~\cite{dodds2009b}.

LIWC, along with all word data sets, should not be 
considered an unimpeachable ``gold standard''; language is far too complex
to make such an assured statement.
All word data sets, including
our own, will have limitations.

\section{Comparison to Warriner and Kuperman's data set}

We next contend with a comparison made by Garcia \etal\ between our 
work on English with a similar sized survey
by Warriner and Kuperman (WK)~\cite{warriner2013a,warriner2014a}.
WK generated a merged list of 13,915 English words, the bulk
of which (11,826) are a list of lemmas taken from movie subtitles.
Immediately, we have a mismatch: our word list incorporated 
the 5000 most frequently used words (or tokens) in each of four disparate 
corpora (New York Times, Google Books, music lyrics, and Twitter)
whereas WK's list is mostly lemmas (e.g., ``sing'' but not ``sung'' or
``sang'') taken from one coherent corpus.
Further, each word was scored by 50 participants in our study,
compared with 14--20 for the WK study.

In their Fig.~1B, Garcia \etal\ show histograms
for the two word lists, which seem to indicate more negative words 
in the WK list and a higher median word happiness for our word list.
But such a comparison is unsound: the words behind each histogram
are not the same and word frequency is not being controlled for.
The two histograms cannot be sensibly compared,
and we can discard Garcia \etal's finding that the median level of average word happiness 
$\havgfn$ for our full data set is 0.28 above the median level for the WK data set.

Nevertheless, Garcia \etal\ do appropriately compare the shared subset of words found
in both data sets, finding a much smaller difference between
median values of $\havgfn$ of 0.07.
They then suggest that our use of cartoon faces
to indicate the 1 to 9 scale of happiness responses
induces a positive bias in respondents' choices, referencing a study that found
a non-smiling face to be slightly negative~\cite{lee2008a}.
Their claim lacks foundation for several reasons.

\begin{figure*}
  \centering
  \includegraphics[width=\textwidth]{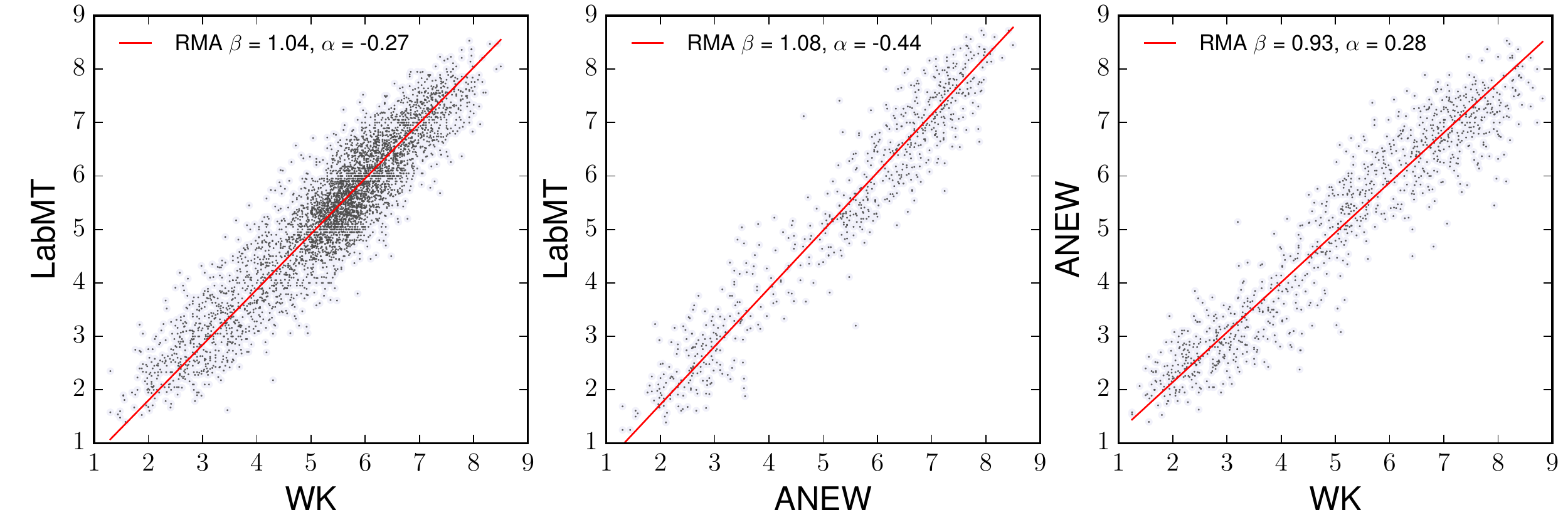}
  \caption{
    Comparison of word ratings for three studies for
    overlapping words:
    labMT~\cite{dodds2011e,dodds2015a_onlineappendices},
    ANEW~\cite{bradley1999a},
    and Warriner and Kuperman~\cite{warriner2013a}.
    Reduced major axis regression~\cite{rayner1985a} yield
    the fits $\havgfn' = \beta \havgfn + \alpha$.
  }
  \label{fig:mhl-reply.WK-comp}
\end{figure*}

First, WK employed a reverse 9 point scale,  
with 1 = happy and 9 = unhappy, flipping the scores after
completing the surveys (also used in~\cite{bradley1999a}).
We have no objection to WK's approach but evidently
this further complicates any comparison between the
two studies. Indeed, one might reasonably 
hypothesize that flipping the direction of the ratings 
could be the sole cause of the minor discrepancy
between the words scored by both studies.

Second, we gave all participants clear written instructions that 5 was neutral.
In wanting to generate results that could be compared with
existing work, we followed the design of Bradley and Lang in their ANEW
study~\cite{bradley1999a}, who used both cartoon figures in their
self-assessment manikins 
and written (spoken for ANEW) instructions;
we departed only in orienting positive to the right.
(As have many others, Garcia \etal\ have used the ANEW study
in their research~\cite{garcia2012a,bradley1999a}.)
WK take pains to compare their scores with the ANEW study (which they
use in part as a control) and other studies, finding their results are ``roughly equivalent'' (p.~6).
And as we noted in~\cite{dodds2015a_onlineappendices}, basic function 
words which are expected to be neutral
such as ``the'' and ``of'' were appropriately scored as such,
indicating that the survey mechanism we used 
was not adding a simple positive shift.

Third, given the nature of language and surveys and changing demographics
online, an exact match for the medians would be a remarkable achievement.
The agreement between the labMT (our English word set~\cite{dodds2011e,kloumann2012a,dodds2015a_onlineappendices})
and WK is still a strong one, 
and we show
scatter plots for the matching word happiness scores in
Fig.~\ref{fig:mhl-reply.WK-comp}
for labMT, ANEW, and WK.
Visually, we see the three studies are sympathetic with each
other, particularly when we acknowledge the typical standard
deviation for the scores of individual words (on the order of 1 to 2).
We used Reduced (or Standard) Major Axis regression~\cite{rayner1985a}
to obtain the fits shown in Fig.~\ref{fig:mhl-reply.WK-comp},
$\havgfn' = \beta \havgfn + \alpha$.
We see that ANEW's scores grow slightly faster than that of both
labMT and WK ($\beta$ = 1.08 and 1.07) and
WK similarly relates to labMT ($\beta = 1.04$).

Fourth, and rather finally, according to the argument of Garcia \etal\ regarding faces, 
the median for ANEW should be higher than that of WK 
(noting again that they used the same happy to sad directionality), 
yet we see the \textit{opposite} (5.29 versus 5.44).
Moreover, comparing medians alone is insufficient---our regression 
analysis shows that, for the words they have in common, 
WK appears more emotionally biased than labMT with $\beta$ = 1.04.
We note that a much richer comparison could be carried out at the level of
individual ratings, but this is far too detailed for the present response.

\section{Dependence of positivity on frequency of usage}
\label{sec:mhl-liwc.freqdepend}

\begin{figure*}
  \centering
  \includegraphics[width=\textwidth]{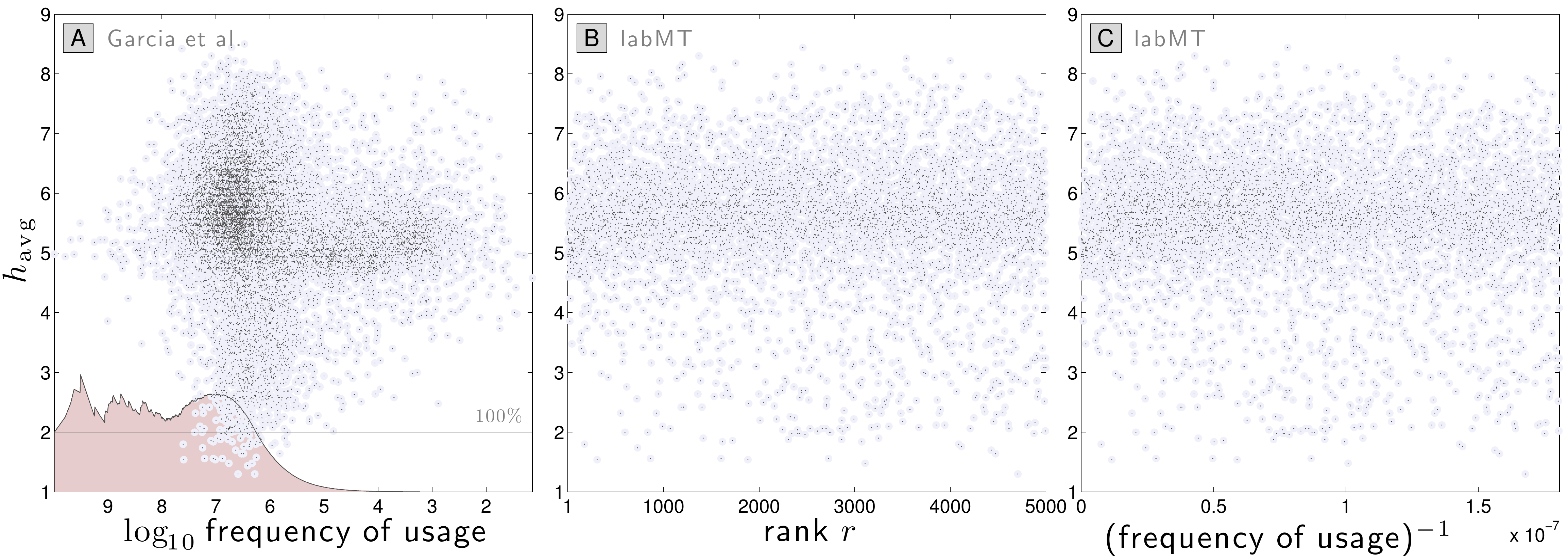}
  \caption{
    \textbf{A.}
    Scatterplot of $\havgfn$ as a function of word usage frequency
    for the English Google Books word list generated by Garcia \etal.
    Uncontrolled subsampling of lower frequency words yields 
    a lexicon that is not statistically representative of any
    natural language corpus.
    The lower curve provides a coarse estimate of cumulative lexicon
    coverage as a function of usage frequency $f$
    using Zipf's law $ f_r \sim f_1 r^{-1} $ inverted
    as $r \sim f_1/f_r$.
    The rapid drop off begins at around rank 5000, the
    involved lexicon size for Google Books in labMT~\cite{dodds2015a_onlineappendices,dodds2011e}.
    \textbf{B.}
    Scatterplot of $\havgfn$ as a function of rank $r$ 
    for the 5000 words for Google Books contributing to labMT,
    the basis of our jellyfish plots~\cite{dodds2015a_onlineappendices}.
    \textbf{C.}
    The same data as in \textbf{B} but now plotted against the
    inverse of usage frequency.
    The approximate adherence to Zipf's law $f \sim r^{-1}$ means
    there is no substantive loss of information if regression is performed
    on the correct transformation of frequency.
Linear regression fits for the first two scatterplots are
    $\havgfn \simeq 0.089 \log_{10} f + 4.85$
    and 
    $\havgfn \simeq -3.04$$\times$$10^{-5} r + 5.62$ (as reported
    in~\cite{dodds2015a_onlineappendices}).
    Note difference in signs, and the far weaker trend
    for the statistically appropriate regression against rank in \textbf{B}.
    Pearson correlation coefficients:
    +0.105, -0.042, and -0.043 
    with $p$-values 
    6.15$\times$$10^{-26}$,
    3.03$\times$$10^{-3}$,
    and
    2.57$\times$$10^{-3}$.
    Spearman correlation coefficients:
    +0.201, -0.013, and -0.013 
    with $p$-values 
    6.37$\times$$10^{-92}$,
    0.350,
    and
    0.350
    (\textbf{B} and \textbf{C} must match).
    The Spearman analysis indicates
    that an assumption of a non-monotonic
    relationship between $\havgfn$ and rank $r$ is well supported.
  }
  \label{fig:mhl-reply.jellyfish}
\end{figure*}

We turn now to Garcia \etal's central claim: that we claimed to find that a positivity
bias is \textit{independent} of word frequency across 10 languages.
In fact, we instead variously stated that a positivity bias is ``strongly'' 
and ``largely'' independent of frequency, and we explored the minor
departures from pure independence in detail for all 24 corpora
across 10 languages (see~\cite{dodds2015a_onlineappendices} and the paper's online appendices).

Garcia \etal\ write that our paper specifically
conflicts with two previous works, their own~\cite{garcia2012a}
and that of Warriner and Kuperman~\cite{warriner2014a}.
We are able to dismiss~\cite{garcia2012a} due to it being founded on a misapplication
of an information theoretic formula by Piantatosi et al.~\cite{piantadosi2011a},
and which we demonstrate elsewhere~\cite{williams2015b}.

Setting aside this misrepresentation,
Garcia \etal's issue with our work becomes to what degree frequency
independence is followed,
and they provide an alternative analysis
of how positivity behaves with usage frequency.
Whereas we performed the regression 
$\havgfn = \alpha r + \beta$ where $r$ is rank,
they claim
$\havgfn = \alpha \log_{10} f + \beta$ is more appropriate.
Once again, we stand by our own principled analysis for the following
reasons.

\textbf{1. Mismatch of scored word list and word list with frequency:}
In attempting to say anything about a given quality of words
as it relates to usage frequency within a specific corpora, a complete census of
words by frequency must be on hand.
Garcia \etal\ have taken our merged word lists for each language
and applied them to data sets for which they do not necessarily fit.
Problematically, their word lists do not contain ranks, and consequently there are words
missing in uncontrolled ways from the data they perform regression on.
For the example of English, our 
10,222 words will (likely) match the most
common words in any sufficiently large English corpus.
But the matching becomes more peculiar the rarer the word,
and the inclusion of Twitter in our word list means
Garcia \etal\ have found ``lolz'', ``bieber'' and ``tweeps'' in 
Google Books.
In Fig.~\ref{fig:mhl-reply.jellyfish}A, we plot average happiness
as a function of frequency of usage for the word list they created
from Google Books.
The scatter plot is clearly unsuitable for linear regression.
We show an estimate of cumulative coverage at the bottom (see caption),
which crashes soon after reaching 5000 words.

\textbf{2. Rank is an appropriate variable to regress happiness (or
  any word quality) against:}
Garcia \etal\ state that regression against frequency $f$
is a better choice because information is lost in moving to rank $r$.
However, the general adherence of natural language
to Zipf's law, $f \sim r^{-1}$, provides an immediate
counterargument~\cite{zipf1949a},
even acknowledging the possibility of a scaling break~\cite{williams2015b}.
Because word usage frequency is so variable, great care must be taken
with any analysis.
As we show for the case of English Google Books in Fig.~\ref{fig:mhl-reply.jellyfish}A, 
regression on $\log_{10} f$ will be gravely compromised by the 
increasing preponderance of words at lower frequencies (a common
issue with measuring power-law slopes), and, based on even the words
for which coverage is reasonable, it would evidently be
in poor judgment to extrapolate from any linear fit across frequencies.
By contrast, Fig.~\ref{fig:mhl-reply.jellyfish}B shows
how usage rank is perfectly suited for regression, and
is the basis for the ``jellyfish'' plots we provided
in Fig.~3 of~\cite{dodds2015a_onlineappendices} and in the paper's online appendices.
Our jellyfish plots make the general conformity 
to a rough (we do not claim ``physical-law'' strict) scale 
independence abundantly clear.
By using rank, we are able to perform a much finer analysis than
Garcia et al. propose, and we show in all corpora that the deciles
for a sliding window of 375 ranks changes at most rather slowly.
Finally, in Fig.~\ref{fig:mhl-reply.jellyfish}C, we present how
$h_{\textrm{avg}}$ behaves as a function of $1/f$, 
illustrating both the error of choosing $\log_{10} f$
and that our results will be essentially unchanged if we regress
against $1/f$.

In closing, we emphasize that minor deviations from frequency independence 
remain a secondary aspect of our observation that the Polyanna Hypothesis holds
for a diverse set of languages, and are wholly irrelevant for the instrumental aspect of our work
in creating text-based hedonometric tools.

\acknowledgments
CMD was supported by NSF grant DMS-0940271;
PSD was supported by NSF CAREER Award \#0846668.

\clearpage

\newwrite\tempfile
\immediate\openout\tempfile=startsupp.txt
\immediate\write\tempfile{\thepage}
\immediate\closeout\tempfile

\setcounter{page}{1}
\renewcommand{\thepage}{S\arabic{page}}
\renewcommand{\thefigure}{S\arabic{figure}}
\renewcommand{\thetable}{S\arabic{table}}
\setcounter{figure}{0}
\setcounter{table}{0}

\end{document}